\documentclass{Interspeech2024}
\usepackage{comment}
\usepackage{svg}
\usepackage{bbding}
\interspeechcameraready

\title{MMSD-Net: Towards Multi-modal Stuttering Detection}

\name[affiliation={1}]{Liangyu}{Nie}
\name[affiliation={2}]{Sudarsana Reddy}{Kadiri}
\name[affiliation={3}]{Ruchit}{Agrawal}

\address{
  $^1$University of Texas at Dallas, USA\\
  $^2$University of Southern California, USA \\
  $^3$University of Birmingham, UAE}
\email{liangyu.nie@utdallas.edu,skadiri@usc.edu,r.r.agrawal@bham.ac.uk}


\newcommand{\red}[1]{\textcolor{red}{#1}}

\begin{document}

\maketitle
\begin{abstract}
Stuttering is a common speech impediment that is caused by irregular disruptions in speech production, affecting over 70 million people across the world. Standard automatic speech processing tools do not take speech ailments into account and are thereby not able to generate meaningful results when presented with stuttered speech as input. The automatic detection of stuttering is an integral step towards building efficient, context-aware speech processing systems. While previous approaches explore both statistical and neural approaches for stuttering detection, all of these methods are uni-modal in nature. This paper presents MMSD-Net, the first multi-modal neural framework for stuttering detection. Experiments and results demonstrate that incorporating the visual signal significantly aids stuttering detection, and our model yields an improvement of 2-17\% in the F1-score over existing state-of-the-art uni-modal approaches.%We conduct experiments on publicly available datasets and compare the results obtained by MMSD-Net with existing baselines. Results
\end{abstract}

\section{Introduction}
Recent advancements in machine learning have enabled a wide range of AI applications across a myriad of sectors such as government, industry, healthcare and transportation. The automatic recognition and transcription of speech is an integral task in machine learning, which enables users to interact seamlessly with machines, lending itself applicable to a variety of tasks such as automatic translation of speech, subtitling of multimedia content, audio signal analysis and editing, and so on. In particular, the massive developments in speech processing have given rise to an unprecedented surge of virtual/digital assistants such as Siri and Alexa, used by millions of users across the globe.  However, these modern speech processing tools are not perfect, and are unable to deal with a range of speech impediments.% such as slurred speech and stuttered speech. 

\par Stuttered speech is one such common speech impairment, wherein the production of fluent speech is hindered by a malfunctioning of the central nervous system of the afflicted person.
This makes it highly challenging for people who stutter to access popular speech recognition tools such as Siri and Alexa. As an example, Apple's Siri reported an accuracy ranging from 18.2-73\% when given stuttered speech as input, compared to a high accuracy of 92\% when presented with normal speech as input \cite{mullin2016siri}, which makes it impractical to be used by people who stutter. Stuttering can manifest itself in various types of disfluencies, such as sound repetition, part-word repetition, word repetition, phrase repetition, revision, interjection, prolongation and block \cite{prasse2008stuttering}. The lack of robust speech recognition of stuttered speech is an unfair consequence for a significant portion of the world population, with over 70 million people being affected by this condition \cite{yairi2013epidemiology}. 
\par Automatic stuttering detection is an integral step towards mitigating this limitation of existing speech processing tools. While a number of approaches have been proposed in the recent years for stuttered speech detection, these are either audio-based or text-based, and therefore \textit{uni-modal} in nature.
%primarily rely on the audio signal and analyse the audio input to identify the portions containing stuttering, and are thereby \textit{uni-modal} in nature. 
The application of multi-modal neural networks for stuttering detection remains unexplored.  This paper aims to bridge this gap and presents the first multi-modal neural network framework for stuttered speech detection.  The primary motivation behind our proposed multi-modal approach is that cues to detect stuttering could be found not only in the audio signal, but also on the speakers' faces. Our hypothesis is that these visual signals contain relevant information pertitent to the SD task. 
%\par Recent methods have explored the usage of neural networks for the identification and removal of stuttering; however, these primarily rely on the audio signal and analyse the audio input to identify the portions containing stuttering. While previous approaches on stuttering detection focus on \textit{uni-modal} methods (audio-based or text-based), we propose a \textit{multi-modal} framework that leverages video data in addition to audio data for stuttering detection.
We validate our hypothesis by conducting experiments across a range of settings and demonstrating results on publicly available datasets. \\ \\
%Templates are provided on the conference website for Microsoft Word\textregistered, and \LaTeX. We strongly recommend \LaTeX\xspace which can be used conveniently in a web browser on \url{overleaf.com} where this template is available in the Template Gallery.
The primary contributions of this paper are summarized below: 
\begin{itemize}
    \item We present MMSD-Net, the first multi-modal neural network for automatic stuttering detection.
    \item Our proposed architecture effectively integrates audio, video, and language data using a novel multi-modal fusion mechanism, which enhances %. The custom attention mechanism with semantic and complementary modules enhances 
    feature fusion for superior multi-modal task performance.
    \item We conduct experimental studies using publicly available datasets and present extensive comparisons of the results obtained by our model against the state-of-the-art uni-modal methods for stuttering detection.
    \item We demonstrate that MMSD-Net outperforms state-of-the-art uni-modal methods by 2-17 \% on F1-score.
    \item We release the code for pre-processing, post-processing as well as the neural network models publicly to ensure reproducibility of our research.
\end{itemize}

\section{Related Work}
The automatic detection of stuttering has been approached through various methodologies, primarily divided into statistical approaches and deep learning approaches. Statistical methods rely on feature extraction from speech signals and subsequent classification using machine learning algorithms. These approaches often employ features such as autocorrelation functions, spectral measures, and Mel-frequency cepstral coefficients (MFCCs), utilizing classifiers like support vector machines (SVM), hidden Markov models (HMM), and artificial neural networks (ANN) \cite{howell1995automatic,howell1997development,tan2007application,ravikumar2009approach}. While statistical methods have shown promising results, they often require manual feature engineering and may not generalize well across different datasets due to their dependency on specific feature sets and classifiers. 
%Within statistical approaches, SVMs have emerged as a popular choice for stuttering detection due to their ability to handle high-dimensional data and nonlinear decision boundaries
 For example, SVMs may struggle with datasets containing imbalanced classes or noisy features, limiting their performance in real-world scenarios \cite{cervantes2020comprehensive}. Similarly, HMMs have been successful in modeling temporal sequences in speech, but they may struggle with capturing complex dependencies in stuttered speech, particularly in distinguishing between different types of disfluencies \cite{ravikumar2009approach}.

%The automatic detection of stuttering has been explored traditionally using signal processing approaches, with features such as autocorrelation function, spectral information, envelope parameters and oscillographic and spectrographic parameters  \cite{howell1995automatic}, \cite{howell1997development}.
%Towards the early 2000s, advancements in Machine Learning led to the development of better methods for stutter classification based on Hidden Markov Models (HMMs) and Support Vector Machines (SVMs) using the Mel-frequency cepstral coefficients (MFCCs) as features \cite{tan2007application}, \cite{ravikumar2009approach}. 
\par The advancements in deep learning have led to the exploration of deep neural networks for stutter detection %based on neural networks such as Convolutional Neural Networks (CNNs) and Recurrent Neural Networks (RNNs)
\cite{resnet,villegas2019novel,zayats2016disfluency}. Recent approaches have explored the usage of the perceptron model \cite{villegas2019novel} and autoregressive models such as Long-Short Term Memory networks (LSTMs) \cite{zayats2016disfluency,santoso2019classification}. While the former employs LSTMs with integer linear programming \cite{georgila2009using}, the latter employs bidirectional LSTMs with attention using the MFCC features. Authors in \cite{resnet} propose a CNN-based model to learn stutter-related features, formulating SD as a binary classification problem. The only input features used in this study are the spectrograms.  Another approach called FluentNet \cite{kourkounakis2021fluentnet} builds upon this method and 
%This paper proposes an acoustic-based deep learning model (residual network + Bi-LSTM) for classifying different stutter types without relying on language models, achieving a significant improvement (10.03\% miss rate) over prior methods.
%Building upon this method, \cite{kourkounakis2021fluentnet} proposes FluentNet, and 
explores the use of a residual network along with an LSTM network to learn frame-level representations. %, and develop a synthetic stuttered speech dataset. 
However, these methods only consider a small subset of disfluent speakers in their studies, and are not tested exhaustively on their ability to generalize well to a variety of stuttered speakers. 
\par With the advent of the Transformer model \cite{vaswani2017attention,chen2020controllable} propose using a controllable time-delay transformer architecture, but only for Chinese data. A similar approach proposed recently, called  StutterNet \cite{sheikh2021stutternet}, employs a time delay neural network and formulates the stutter detection task as a multi-class classification problem. However, this method only consider limited disfluent behaviours (blocks, repetition, and prolongation) in addition to fluent speech segments. 
A major challenge to train deep neural networks is the availability of large-scale annotated datasets. To address this problem, \cite{lea2021sep} curate a large-scale dataset for stuttering detection, called SEP-28k. They also present experiments using the ConvLSTM model and demonstrate results on the FluencyBank as well as SEP-28k datasets. More recently, \cite{bayerl2022detecting} employ the Wav2Vec model for stuttering detection and explore self-supervised learning for this task. 
%\par \textcolor{blue}{add paragraph for FluentSpeech}
The closest to our work is the method called \textit{FluentSpeech} proposed by \cite{jiang2023fluentspeech}. However, it is uni-modal and does not leverage the visual signal. Additionally, it considers pauses, non-lexical vocalisations and interjections such as \textit{so, hmmm, umm, like} in speech as stuttered segments \cite{roberts2009disfluencies}. We disagree from their method in that such normal disfluencies could correspond to useful pauses in speech wherein the speaker can plan their upcoming discourse; and are notably different from \textit{stuttering}, which is a neuro-developmental speech disorder that is characterized by core behaviour and corresponds to abnormally persistent stoppages in normal speech, often accompanied by unusual behaviours such as quick eye blinks, lip tremors and nodding of head \cite{riva2008phenomenology}.

\par It must be noted that while there have been several approaches proposed for stuttering detection,  these rely on classical machine learning (typically on the audio signal) or uni-modal deep learning methods. 
%all the approaches proposed for stuttering detection primarily rely on the audio signal and analyse the audio input to identify the portions containing stuttering, and are thereby uni-modal in nature.
%While there have been several approaches proposed for stuttering detection, most of these rely on classical machine learning or uni-modal deep learning methods. 
The application of multi-modal deep learning remains unexplored for this task. We bridge this gap and present MMSD-Net, the first multi-modal deep learning method for stutter event detection.
%\textcolor{red}{Sudarsana: Can you edit the previous 2-3 paragraphs to compress it a bit and add any limitations if you find?}. 

\section{Proposed Methodology}
\begin{figure}[th!]
  \centering
  \vspace{-0.4cm}
  \includegraphics[width=\columnwidth,height=4cm,trim={1cm, 0cm, 1cm, 1cm},clip]{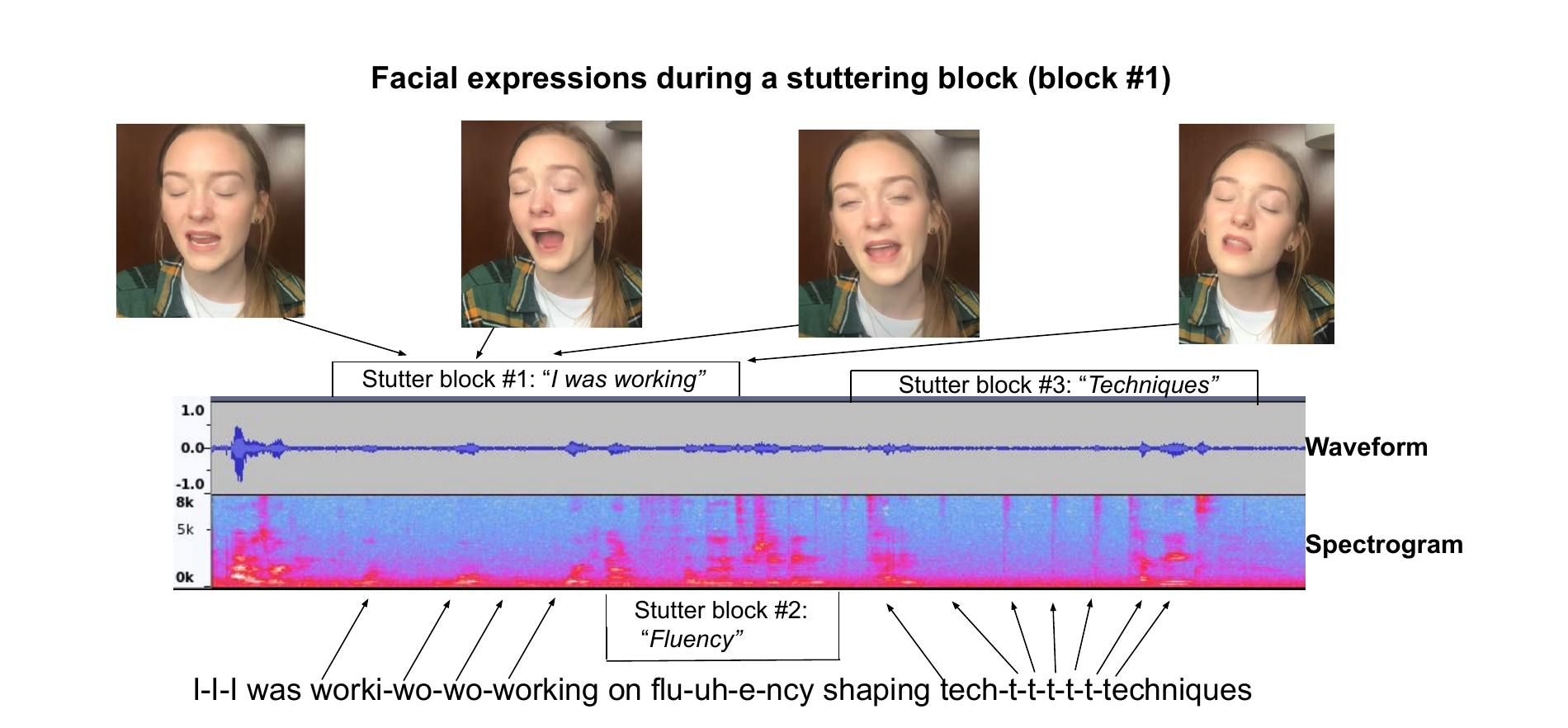}
  \caption{Illustration of multi-modal cues during stuttering. \\ Sentence: "I was working on fluency shaping techniques."}
  \label{fig:motivation}
\end{figure}
This section presents a detailed overview of MMSD-Net. Technically, stutter presents itself as an audiovisual problem.
%however previous approaches on stuttering either employ signal processing on the audio signal or use data-driven methods using audio-only or text-only inputs. 
Figure \ref{fig:motivation} demonstrates a data sample from the \textit{Adults Who Stutter} dataset \cite{ratner2018fluency}, depicting a snippet of the video, and the corresponding waveform, spectrogram and textual transcription respectively. The primary motivation of our method is that cues for stuttering can be found from the visual signal as well as the audio signal, as illustrated in Figure \ref{fig:motivation}.  It can be observed that facial expressions change during stuttering and can thereby provide important contributions for the automatic stutter detection task. To this end, we employ the visual signal in addition to the audio signal, and present MMSD-Net, the first multi-modal neural approach towards stuttering detection.\\ \\
\textbf{\underline{Model architecture:}} Figure 2 presents an overview of our model architecture. MMSD-Net comprises three primary modules, described in the subsequent subsections:

\begin{figure*}[th]
  \centering
  \includegraphics[width=4in]{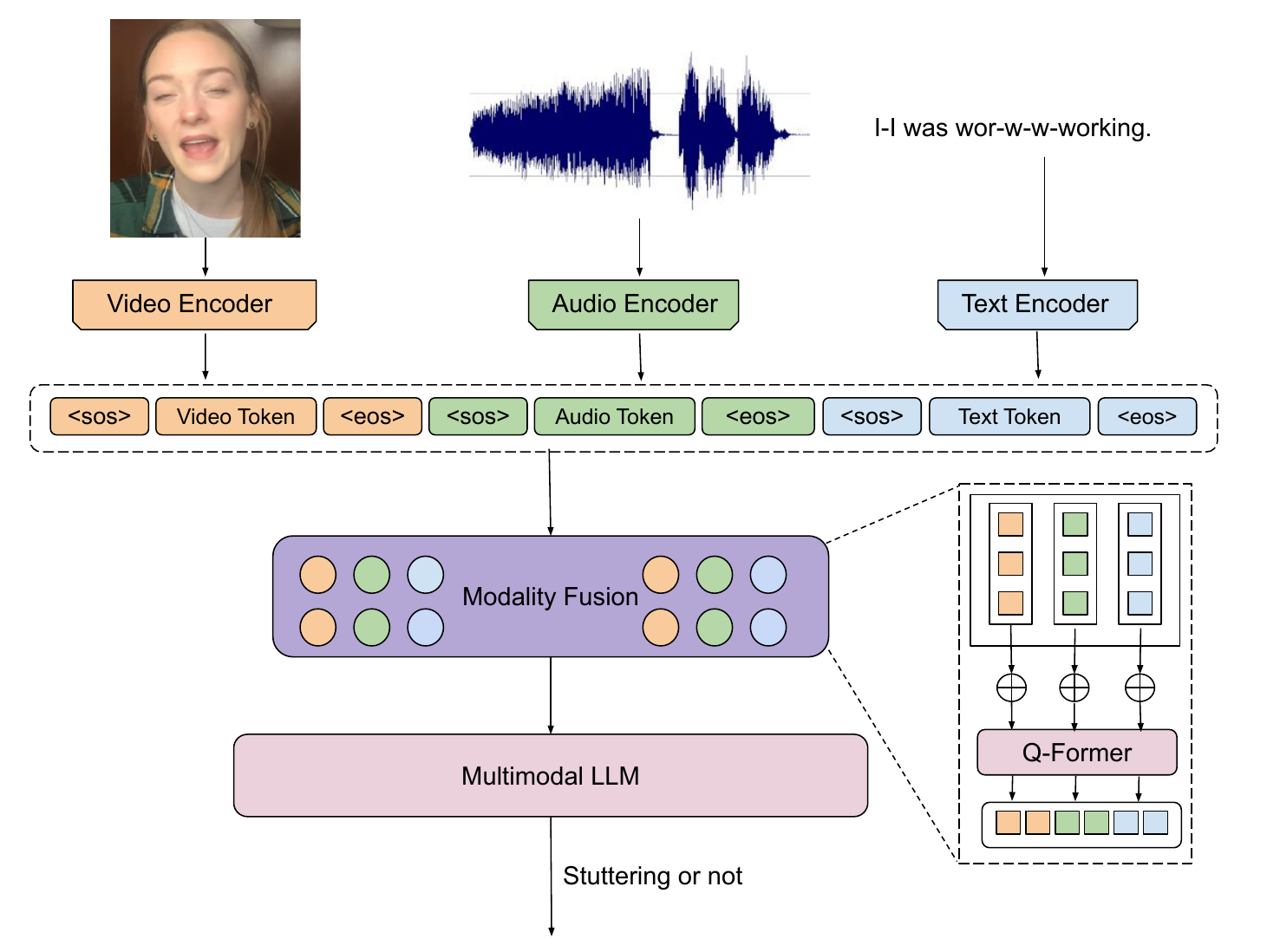}
  \vspace{-0.3cm}
  \caption{Detailed architecture of our proposed model MMSD-Net}
  \label{fig:architecture}
\end{figure*}

%\textcolor{red}{Diagram of stuttering (snippets of video, waveform and spectrogram) to be added by Sudarsana}.  
%\subsection{Model Architecture}
%This section provides an overview of MMSD-Net. 

%It starts with an overview of the model's architecture and then moves on to an elaborate examination of each component within MMSD-Net. The section concludes with a thorough discussion on the training methodology of MMSD-Net.
%\begin{enumerate}
    \subsection{Multi-encoder module}
    Current Multimodal Language Models are mainly geared towards understanding video and text-based content. MMSD-Net incorporates specialized modality encoders to process not just video and text, but also auditory data. These encoders are designed to extract the most relevant features from each modality.
This upgrade significantly enhances MMSD-Net's ability to process and interpret information across various modalities efficiently. Unlike traditional models that might rely on Convolutional Neural Networks (CNNs), MMSD-Net leverages three transformer encoders for video, audio and text data respectively. Transformers excel at capturing long-range dependencies within sequences, making them well-suited for  stuttering detection. %where the order and relationships between videos are crucial. 
By capturing these nuanced details, MMSD-Net gains a richer understanding of the input data corresponding to  different modalities, allowing for more comprehensive reasoning. %alongside the information from video, audio and text.
The specialized encoders process the video (\textbf{v}), audio (\textbf{a}) and textual (\textbf{t}) inputs as follows:

\begin{equation}
  \mathbf{h}_v = \text{Model}(\mathbf{v}) \in \mathbb{R}^{d_v}
\end{equation}
\begin{equation}
  \mathbf{h}_a = \text{Model}(\mathbf{a}) \in \mathbb{R}^{d_a} 
\end{equation}
\begin{equation}
  \mathbf{h}_t = \text{Model}(\mathbf{t}) \in \mathbb{R}^{d_t}
\end{equation}
%$\mathbf{h}_v \in \mathbb{R}^{l_v \times d_h}$ and $\mathbf{h}_a \in \mathbb{R}^{l_a \times d_h}$ and $\mathbf{h}_t \in \mathbb{R}^{l_t \times d_h}$, are the video, audio, and textual features respectively, and $d_h$ is the dimension of modality-specific features.
\par where $\mathbf{h}_v$, $\mathbf{h}_a$, and $\mathbf{h}_t$ represent the extracted features for video, and audio, respectively. $d_v$, $d_a$, and $d_t$ denote the dimensionality of the features for each modality. \\
\par To reduce computational costs and minimize the number of tokens in the prefix, we employ a 1-D convolutional layer to compress the length of the multi-modal features to a smaller value. Subsequently, a linear layer is employed to adjust the hidden size of the features, fusing it with the size of the MLMs' embeddings as follows:

\begin{equation}
\label{eq:6}
 \mathbf{h}'_v = \text{Linear}(\text{Conv1D}(\mathbf{h}_v))
\end{equation}
\begin{equation}
\label{eq:7}
  \mathbf{h}'_a = \text{Linear}(\text{Conv1D}(\mathbf{h}_a))
\end{equation}
 \begin{equation}
 \label{eq:8}
    \mathbf{h}'_t = \text{Linear}(\text{Conv1D}(\mathbf{h}_t)) \\
 \end{equation} \\ 
where $\mathbf{h}'_v, \mathbf{h}'_a, \mathbf{h}'_t$ are the transformed features with a fixed length of $L'$ and an embedding dimension of $d_e$. The value of $L'$ is significantly smaller than $L_v$, $L_a$, and $L_t$, while $d_e$ corresponds to the dimensionality of the embedding matrix $\mathbf{E} \in \mathbb{R}^{v \times d_e}$ associated with the MMSD-Net.
    \subsection{Multimodality Fusion Module}
    Modality encoders are often trained independently, which can result in differences in the representations they produce. Therefore, it is essential to merge these distinct representations into a unified space. %This section describes the method we use to integrate these representations.
%    \textbf{Multimodal Fusion:}\\
%The fusion strategy is designed to 
We design our fusion strategy based on the Q-Former \cite{Q-former}, and modify it to efficiently integrate video and audio features with textual features, thereby facilitating more rapid adaptation. In this context, we designate the video features derived from our visual modality encoder by \(h_v \in \mathbb{R}^{L_v \times d_v}\), and the audio features extracted from the audio modality encoder are denoted as \(h_a \in \mathbb{R}^{L_a \times d_a}\).% The methodology for fusioning these different modalities is described as follows:\\ 
\par In order to fuse the distinct representations learnt separately by the individual modality encoders, we consider the transformed visual and audio modality representations obtained in Equation \ref{eq:6}, \ref{eq:7} and \ref{eq:8} as the soft tokens of our MLM module. The visual and audio representations are fused with the textual embedding space using the attention mechanism from Equation 1, as follows:
\begin{equation}
  \mathbf{h}^a = \text{Attn}(\mathbf{h}', \mathbf{E}, \mathbf{E})
\end{equation}
where $\mathbf{h}'$ is the modality representation obtained in Equations \ref{eq:6}, \ref{eq:7}, and \ref{eq:8}, \textbf{E} is the embedding matrix as defined in Section 3.1, and $\mathbf{h}^a$ is the corresponding fused representation, specifically $\mathbf{h}^a_v$, $\mathbf{h}^a_a$, and $\mathbf{h}^a_t$. After this fusion operation facilitated by the attention mechanism, the MMSD can seamlessly process the representations from various modalities.\\ \\
%\textbf{Integration:}\\
%The integration of fusioned modality representations into the instruction can be achieved effortlessly through the concatenation operation. 
In order to integrate the fused modality representations with the instruction information, we employ the concatenation operation. Given the fused modality representations, we define the integration as follows
\begin{equation}
  x = [h^a_v:h^a_a:h^a_t:Embed (x_t)]
\end{equation}
where $[:]$ represents the concatenation operation, $x$ represents the multi-modal instruction,
$x_t$ represents the sequence of tokens in the textual instruction, and $Embed$ ($x_t$) represents the sequence of embeddings of $x_t$.
    \subsection{MLM module}
    Multimodal Language Models (MLM) \cite{mlm} 
    have demonstrated exceptional aptitude in comprehending and executing human directives. In particular, cross-modal approaches have demonstrated improved performance for audio synchronization tasks \cite{agrawal2021structure}, \cite{agrawal2021convolutional}. In MMSD-Net, we utilize pretrained MLMs as the core modules, establishing the basis of MMSD-Net's functionality. The pretrained MLM network processes three primary inputs: 
   % \textbf{Multihead Self Attention: }\\The Transformer model incorporates a critical element known as scaled dot-product attention, as outlined in \cite{attention}. 
    %This mechanism processes three primary inputs: 
    the query vector \(Q \in \mathbb{R}^{n_q \times d_q}\), the key vector \(K \in \mathbb{R}^{n_k \times d_k}\), and the value vector \(V \in \mathbb{R}^{n_v \times d_v}\). Through the scaled dot-product attention, the model evaluates attention scores by comparing each query in \(Q\) with all keys in \(K\). These scores are then utilized to refine the query representations by generating a weighted sum of the values in \(V\). The operation is mathematically represented as:
    \[Attention(Q, K, V) = softmax(\frac{QK^T}{\sqrt{d_k}})V\]
        Here, \(d_k\) denotes the size of the key and query vectors, while \(n_q\) and \(n_k\) represent the counts of queries and keys, respectively.
    \[MultiHead(Q, K, V) = Concat(head_1, ..., head_h)W^O\]
    We use stacked multi-head attention as well as positional encodings to model the complete sequential information encoded by the inputs. The decoder employs masked multi-head attention followed by softmax normalization for binary classification. The positional encodings are added to the input as well as output embeddings, enabling the model to capture the sequentiality of the input sentence without having recurrence. The encodings are computed from the position ($pos$) and the dimension ($i$) as follows:
\begin{equation}
PE_{(pos, 2i)} = sin(pos/10000^{(2i/d_{model})})
\end{equation}
\begin{equation}
PE_{(pos, 2i + 1)} = cos(pos/10000^{(2i/d_{model})})
\end{equation}
where $PE$ stands for positional encodings and $d_{model}$ is the dimensionality of the vectors resulting from the embeddings learned from the input and output tokens. 

\section{Experiments and Results}
%\subsection{Experimental setup}
%This sections describes the experiments carried out in this paper.
We conduct experiments for the detection of stuttered speech using our proposed model MMSD-Net, and compare the results obtained by our model with four state-of-the art baseline models, i.e. FluentSpeech \cite{speechFluent}, ResNet+BiLSTM \cite{resnet}, ConvLSTM \cite{lea2021sep} and StutterNet \cite{sheikh2021stutternet}. 
%models in the same settings to have a
%fair comparison.
\subsection{Datasets}
This study uses four publicly available datasets (1 audio-modality and 3 audio-visual modalities) for the experiments. The details of modalities present in each dataset, along with the amount of data used for training and testing in each of our experiments, are described in Table \ref{table:datasets}.

\begin{table}[th!]
\caption{Details of datasets used for our experiments.}
\vspace{-0.1cm}
\scalebox{0.95}{
\begin{tabular}[t]{p{2.2cm} p{1cm} p{1.1cm} p{1.1cm} p{1.1cm}}
\toprule
\textbf{Dataset Name}&\textbf{Modality}  & \textbf{Content}&\textbf{Training set} &\textbf{Testing \newline set} \\
\midrule
Sept28K \cite{lea2021sep} &Audio&Normal Speech&28,000&0 \\
FluencyBank \cite{ratner2018fluency}&Video, \newline Audio &Normal\newline+stuttered speech&52,000&0\\
Adults Who \newline Stutter \cite{ratner2018fluency} & Video, \newline Audio &Stuttered Speech& 200&500\\
SpeakingFaces \cite{abdrakhmanova2021speakingfaces}&Video, \newline Audio&Normal speech & 200&500\\
\midrule
\end{tabular}}
\captionsetup{justification=centering}
\label{table:datasets}
\end{table}

\subsection{Experimental Setup and Hyperparameters}
For training efficiency, we leverage LoRA \cite{lora} for optimization on 4 Nvidia A100 GPUs. Each GPU handles a batch size of 4, with gradients accumulated for 5 steps before updating the model. The training process lasts for 10 epochs, employing a cosine learning rate scheduler with an initial learning rate of $5 \times 10^{-10}$ and a warmup ratio of 0.02. To promote efficiency, FP16 precision is used for both training and inference. The maximum sequence length is capped at 512 tokens.
\subsection{Results and Discussion}
Table~\ref{results} gives the results (in terms of precision, recall and F1-score) of the proposed MMSD-Net method along with the four baseline methods. The results are presented for the dataset obtained by combining the FluencyBank \cite{ratner2018fluency}, Sept28K \cite{lea2021sep} and Adults Who Stutter \cite{ratner2018fluency} datasets as described in Table~\ref{table:datasets}. In addition to model performance, the comparison of modalities used by each method is also presented in the Table~\ref{results}. Among the four baseline methods, StutterNet \cite{sheikh2021stutternet} achieved better results compared to other three methods (FluentSpeech \cite{speechFluent}, ResNet+BiLSTM \cite{resnet} and ConvLSTM \cite{lea2021sep}), and ResNet+BiLSTM \cite{resnet} achieved the lowest scores among the baseline methods. While both StutterNet \cite{sheikh2021stutternet} and ResNet+BiLSTM \cite{resnet} are specifically trained on audio data, StutterNet achieves superior performance using MFCC features extracted from the audio samples. This suggests that MFCC features are more suitable for stutter detection compared to the features employed by ResNet+BiLSTM, i.e. the spectrograms.
\par Our proposed MMSD-Net outperforms all other methods in terms of precision, recall, and F1-score, achieving the highest scores across all metrics, which demonstrates its superiority in stuttering detection. Our findings demonstrate that the fusion module can effectively combine information from three different modalities.   Quantitatively, the proposed MMSD-Net gave an absolute improvement of 2\% in the F1-score (and 3\% in the Precision) over the best baseline method (StutterNet \cite{sheikh2021stutternet}) and 16\% in the F1-score (and 17\% in the Precision) over ResNet+BiLSTM \cite{resnet}. These results validate our hypothesis that facial expressions serve as an important cues to detect stuttered speech, and employing the visual signal as part of a multi-modal framework improves the performance of automatic stuttering detection.

\begin{table}[ht!]
\caption{Comparison of results obtained by the proposed MMSD-Net with the four baseline methods. Best result highlighted in bold, second best underlined. \textbf{P} = Precision, \textbf{R} = Recall, \textbf{F1} = F1-score.}. 

\begin{tabular}{ p{2.6cm} p{0.6cm} p{0.6cm}  p{0.6cm} p{0.6cm} p{0.6cm} }
% \hline
% \multicolumn{6}{|c|}{Model Comparison} \\
 \toprule
 \textbf{Model Name}& \textbf{Audio} &\textbf{Video} &\textbf{P} & \textbf{R}& \textbf{F1}\\
 \midrule
 FluentSpeech \cite{speechFluent}    &  \Checkmark & \XSolidBrush & 85.73 & 81.82 &83.72\\
 ResNet+BiLSTMs \cite{resnet}   &\Checkmark&   \XSolidBrush & 75.28 & 72.73&73.98\\
 StutterNet \cite{sheikh2021stutternet} & \Checkmark &\XSolidBrush  & \underline{89.41} & \underline{87.10} & \underline{88.23}\\
 ConvLSTM \cite{lea2021sep}   &  \Checkmark    &\XSolidBrush & 82.63 & 78.32 &80.41\\
 \textbf{MMSD-Net} &   \Checkmark   &\Checkmark & \textbf{92.58} & \textbf{87.93}& \textbf{90.19}\\
 \midrule
\end{tabular}
\label{results}
\vspace{0.1cm}
\end{table}
\begin{comment}
\begin{tabular}{ |p{2.2cm}|p{1.5cm}|p{1.5cm}|p{1.5cm}|  }
 \hline
 \multicolumn{3}{|c|}{Result} \\
 \hline
 Model Name& Accuracy & F1\\
 \hline
 FluentSpeech\cite{speechFluent}  & 28.3 & F1 \\
 ResNet+BiLSTMs&   15.2 & F1 \\
 StutterNet &21.74 & F1\\
 ConvLSTM     & 11.9 & F1\\
 MMSD-Net&   31.7   & F1\\
 \hline
\end{tabular}
\end{comment}
\vspace{-0.2cm}
\section{Conclusion}
This study presents MMSD-Net, the first multi-modal neural framework crafted explicitly for stuttered speech detection. We conducted experiments on publicly available datasets and performed studies comparing against four existing uni-modal baselines. Our findings showcase noteworthy improvements in stuttered speech detection accuracy, with enhancements ranging from 2-17\% in F1-score over established baseline models, indicating the effectiveness of our multi-modal approach for stuttering detection. This paper signifies a major stride towards augmenting the efficacy of stuttered speech detection, and highlights the complementarity of multiple modalities for this task. In particular, it demonstrates that the visual signal carries relevant information for the stuttering detection task, and lays the groundwork for the development of further advancements in speech processing tools catered to individuals suffering from speech impediments. %It lays the groundwork for the development of more efficient tools tailored to individuals with speech impediments. 
In the future, we would like to extend our experimentation to larger datasets and conduct a qualitative analysis of the impact of multi-modality on handling various kinds of stuttering.
\bibliographystyle{IEEEtran}
\bibliography{mybib}

\end{document}